\documentclass[aps,prb,showpacs,twocolumn,floatfix]{revtex4}

\usepackage{graphicx}
\usepackage{bm}
\usepackage{amssymb}
\usepackage{amsmath}

\begin{document}

\title{Real space finite difference method for conductance calculations}
\author{Petr A. Khomyakov}
\author{Geert Brocks}
\thanks{Corresponding author}
\email[E-mail: ]{g.brocks@tn.utwente.nl}
\homepage{http://www.tn.utwente.nl/cms} \affiliation{Computational
Materials Science, Faculty of Science and Technology and MESA$^+$
Research Institute, University of Twente, P.O. Box 217, 7500 AE
Enschede, the Netherlands}
\date{\today}

\begin{abstract}
We present a general method for calculating coherent electronic transport in quantum wires
and tunnel junctions. It is based upon a real space high order finite difference
representation of the single particle Hamiltonian and wave functions. Landauer's formula is
used to express the conductance as a scattering problem. Dividing space into a scattering
region and left and right ideal electrode regions, this problem is solved by wave function
matching (WFM) in the boundary zones connecting these regions. The method is tested on a
model tunnel junction and applied to sodium atomic wires. In particular, we show that using a
high order finite difference approximation of the kinetic energy operator leads to a high
accuracy at moderate computational costs.
\end{abstract}


\maketitle

\section{Introduction}\label{intro}
The progress in experimental control on the nanometer scale has enabled studies of electronic
transport in quantum wires of atomic dimensions.\cite{ruitenbeek1} The transport properties
of such systems have to be understood on the basis of their atomic structure. This notion has
generated a large effort in recent years to calculate the conductance of quantum wires from
first principles. Several different approaches have been formulated, which have a common
basis in the Landauer-B\"{u}ttiker approach to express the conductance of a coherent system
in terms of a quantum mechanical scattering problem.\cite{landauer} In such calculations the
quantum wire consists of a scattering region of finite size, sandwiched between two
semi-infinite leads that are considered to be ideal ballistic wires. Semi-empirical
tight-binding models have been exploited to solve this problem.
\cite{todorov,mackinnon,datta1,emberly} Aiming at a better description of the electronic
structure, several current approaches rely upon density functional theory (DFT).

The main differences between these approaches lie in the approximations that are used to
describe the atomic structure of the leads and in the techniques that are used to solve the
scattering problem. In pioneering work, jellium (i.e. free electron) electrodes have been
used to describe the leads and the scattering wave functions have been obtained by a transfer
matrix method\cite{tsukada} or by solving the Lippman-Schwinger equation.\cite{lang,diventra}
A transfer matrix method has also been used taking into account the full atomic structure of
the leads at the DFT level.\cite{choi1,choi2} Alternatively, the conductance can be
calculated using a Green function approach without calculating the scattering wave functions
explicitly.\cite{datta} Several implementations of this approach have been formulated that
use a localized basis set to form a representation of the scattering problem. These
implementations mainly differ in the kind of basis set used, e.g. Gaussian or numerical
atomic orbitals, or wavelets.\cite{xue,taylor,nardelli,palacios,heurich,brandbyge,thygesen}
An embedded Green function approach has been applied using a delocalized basis set of
augmented plane waves.\cite{wortmann}

In this paper we present a technique for solving the scattering problem of a quantum wire
without the use of a basis set. Instead, potentials and wave functions are represented on a
uniform real space grid and differential operators are approximated by a finite difference
approximation (FDA). Previous implementations of this idea have used a simple first order
FDA.\cite{vigneron,gagel,hirose,starikov} In that case the grid has to be relatively fine in
order to obtain sufficiently converged results. This hinders the application to large systems
because of the computational costs involved in using fine grids. However, in ground state
(DFT) electronic structure calculations high order FDA's have been shown to markedly increase
the efficiency of real space grid techniques by enabling the use of coarse
grids.\cite{che1,che2,gygi} In this paper we demonstrate that high order FDA's make it
possible to solve the scattering problem much more efficiently.

The method we propose for calculating the conductance of a quantum wire is based upon wave
function matching (WFM) in the boundary zones connecting the leads and the scattering
region.\cite{ando} Unlike transfer matrix methods, however, it does not require the explicit
calculation of wave functions in the scattering region.\cite{tsukada,choi1,choi2} It does not
require the explicit calculation of Green functions
either,\cite{datta,xue,taylor,nardelli,palacios,heurich,brandbyge,thygesen,wortmann} which
enables us to solve the scattering problem at real, instead of complex,
energies.\cite{greenremark} Our method can be classified as an $O(N)$ technique, since the
computing costs are determined by the size of the scattering region with which they scale
linearly. A related technique that uses a linearized muffin tin orbital basis set, has been
applied to calculate the electronic transport in layered magnetic
materials.\cite{kelly1,kelly2} Although the formalism presented here can be extended to the
non-equilibrium situation, we consider in this paper the linear response regime only.

This paper is organized as follows. In Sec.~\ref{method} the main ingredients of our
computational method are explained, where the computational details can be found in appendix
\ref{details}. The accuracy and convergence properties of the method are verified on model
tunnel junctions in Sec.~\ref{tests}. The application to a more complex system, which
consists of a sodium atomic wire, is discussed in Sec.~\ref{realstuff}. A summary is given in
Sec.~\ref{summary}.

\section{Computational Method}\label{method}
Within the Landauer-B\"{u}ttiker approach the conductance $G$ of a quantum wire is expressed
in terms of the total transmission $T(E)$
\begin{equation}\label{f00}
G=\frac{e^2}{\pi\hbar}T(E),
\end{equation}
assuming spin degeneracy.\cite{landauer} $T(E)$ can be obtained by
solving the quantum mechanical scattering problem at the fixed
energy $E$. Eq.~(\ref{f00}) is valid in the linear response
regime, where $T(E)$ needs to be evaluated at the Fermi energy
$E=E_F$. Our quantum wire is defined as a system consisting of a
finite scattering region that is connected left and right to
semi-infinite leads. The latter are supposed to be `ideal' wires,
which can be described by a periodic potential along the wire
direction. In the scattering region the potential can have any
shape. We consider two cases that can be treated by essentially
the same technique. In the first case the system has a finite
cross-section perpendicular to the wire direction, whereas in the
second case the system is periodic perpendicular to the wire. The
latter case also covers planar interfaces and tunnel junctions.

In order to solve the scattering problem we generalize a method formulated by
Ando.\cite{ando} Here one basically solves a single particle Schr\"odinger equation directly
at a fixed energy $E$ in two steps. In the first step one obtains the modes of the ideal
leads. Subsequently the wave functions for the scattering region are constructed such, that
they are properly matched to the solutions in the leads. We use a real space finite
difference method to represent the Schr\"odinger equation. In the following three subsections
we will introduce this representation and discuss the steps required to solve the scattering
problem.

\subsection{Finite difference approximation}\label{finite}
We start from a single particle equation of the general form
\begin{equation}\label{f0}
\left( E - V ({\bf r}) + \frac{\hbar^{2}}{2 m} \nabla^{2} \right) \Psi ({\bf r}) = 0,
\end{equation}
which represents the Schr\"odinger equation of a single particle in a potential $V$.
Alternatively, within the DFT scheme it represents the Kohn-Sham equation with $V$ the total
effective potential. We put the wave function $\Psi$ and the potential $V$ on a equidistant
grid in real space ${\bf r}=(x_j,y_k,z_l)$, where $x_j=x_0+jh_x,y_k=y_0+kh_y,z_l=z_0+lh_z$
and $h_x,h_y,h_z$ are the grid spacings in $x,y$ and $z$ directions, respectively. Following
Refs.~\onlinecite{che1,che2} we replace the kinetic energy operator in Eq.~(\ref{f0}) by a
high order FDA. For the $x$ part this gives
\begin{equation}\label{f01}
\frac{\partial^2\Psi(x_j,y_k,z_l)}{\partial x^2} \approx
\frac{1}{h_x^2}\sum_{n=-N}^N c_{n} \Psi(x_{j+n},y_k,z_l),
\end{equation}
with similar expressions for the $y$ and $z$ parts. Expressions for the coefficients $c_{n}$
for various values of $N$ are tabulated in Ref.~\onlinecite{che2}. The simplest approximation
in Eq.~(\ref{f01}) ($N=1$, where $c_1=c_{-1}=1$ and $c_0=-2$) reduces Eq.~(\ref{f0}) to the
well-known simple finite difference representation of the Schr\"odinger
equation.\cite{datta,vigneron,gagel,hirose,starikov} However, we will demonstrate that the
scattering problem can be solved much more efficiently using higher order FDA's with
$N=4$-$6$.

\begin{figure}[!tbp]
\includegraphics[width=6.5cm,keepaspectratio=true]{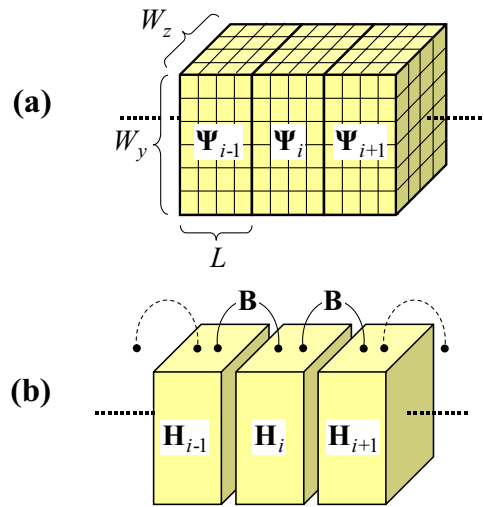}
\caption{(Color online) (a) The system is divided into cells indicated by an index $i$. The
cells have $L \cdot W_y \cdot W_z$ grid points in the $x,y$ and $z$ directions, respectively.
$\mathbf{\Psi}_i$ is the supervector that contains the wave function values on all grid
points in cell $i$. (b) $\mathbf{H}_i$ is the Hamilton matrix connecting grid points within
cell $i$; the $\mathbf{B}$-matrix connects grid points between neighboring cells and is
independent of $i$.} \label{fig:grid}
\end{figure}

In a FDA the Schr\"odinger equation of Eq.~(\ref{f0}) becomes
\begin{eqnarray}\label{f02}
\left(E-V_{j,k,l}\right)\Psi_{j,k,l} + \sum_{n=-N}^N \biglb(
t^{x}_{n} \Psi_{j+n,k,l} + t^{y}_{n} \Psi_{j,k+n,l} + \nonumber
\\
t^{z}_{n} \Psi_{j,k,l+n} \bigrb)  =0,
\end{eqnarray}
where $V,\Psi_{j,k,l}$ is a short-hand notation for
$V,\Psi(x_j,y_k,z_l)$ and
$t^{x,y,z}_{n}=\hbar^{2}/2mh^{2}_{x,y,z}\times c_{n}$. In order to
make a connection to Ando's formalism\cite{ando} we divide the
wire into cells of dimension $a_x \times a_y \times a_z$. The
direction of the wire is given by the $x$-axis. The number of grid
points in a cell is $L=a_x/h_x$ , $W_y=a_y/h_y$, $W_z=a_z/h_z$ for
the $x$, $y$ and $z$ directions respectively. We wish to
distinguish between two different cases. In the first case the
wire has a finite cross-section in the $yz$ plane. In the second
case the wire has an infinite cross-section, but it has a periodic
potential in the $yz$ plane, i.e.
$V_{j,k+W_y,l}=V_{j,k,l+W_z}=V_{j,k,l}$. In both cases the (unit)
cell in the $yz$ plane is described by $W_y \times W_z$ grid
points.

The values $\Psi_{j,k,l}$ where the indices $j,k,l$ correspond to a single cell $i$ are
grouped into a supervector ${\bm \Psi}_{i}$. The idea is shown in Fig.~\ref{fig:grid}. This
supervector has the dimension $N_{\rm rs}=L \cdot W_y \cdot W_z$, which is the total number
of real space grid points in a cell. If we let $i$ denote the position of the cell along the
wire then Eq.~(\ref{f02}) can then be rewritten as
\begin{eqnarray}\label{f3}
(E{\bf I} - {\bf H}_i)  {\bm \Psi}_{i} + {\bf B}{\bm \Psi}_{i - 1}
+ {\bf B^{\dagger}} {\bm \Psi}_{i + 1} = 0,
\end{eqnarray}
for $i=-\infty,\ldots,\infty$. Here ${\bf I}$ is the $N_{\rm rs}
\times N_{\rm rs}$ identity matrix. The matrix elements of the
$N_{\rm rs} \times N_{\rm rs}$ matrices ${\bf H}_i$ and ${\bf B}$
can be derived straightforwardly from Eq.~(\ref{f02}). The
expressions are given in appendix \ref{handb}, both for a wire
that is finite and for a wire that is periodic in the $yz$ plane.
For the latter ${\bf H}_i={\bf H}_i({\bf k}_\parallel)$, where
${\bf k}_\parallel$ is a wave vector in the two-dimensional
Brillouin zone. In the following this notation is suppressed.

Eq.~(\ref{f3}) has the form of a nearest neighbor tight-binding equation, expressed in terms
of vectors/matrices of dimension $N_{\rm rs}$. This form enables us to use Ando's technique
to solve the scattering problem.\cite{ando} Note however that the matrices ${\bf B},{\bf
B^{\dagger}}$ in Eq.~(\ref{f02}) are singular, see Eq.~(\ref{a02}), which requires a
generalization of this technique.

\subsection{Ideal wire}\label{ideal}
An ideal wire is defined by a potential that is periodic in the
direction of the wire, i.e. $V(x_j+a_x,y_k,z_l)=V(x_j,y_k,z_l)$ or
$V_{j+L,k,l}=V_{j,k,l}$. Since the potential is the same in each
cell, the matrix ${\bf H}_i={\bf H}$ in Eq.~(\ref{f3}) is
independent of the cell position $i$. In a periodic system the
vectors in subsequent cells are related by the Bloch condition
\begin{equation}\label{f6}
\lambda {\bm \Psi}_{i} = {\bm \Psi}_{i + 1},
\end{equation}
where $\lambda = e^{i k_x a_x}$ with $k_x$ real for propagating
waves and complex for evanescent (growing or decaying) waves.
Combining Eqs.~(\ref{f3}) and (\ref{f6}) one then obtains the
following generalized eigenvalue problem
\begin{equation}\label{f7}
 \left[ \left( \begin{array}{cccc}
 E{\bf I} - {\bf H}   & {\bf B} \\
 {\bf I} & 0
\end{array} \right)
-\lambda
 \left( \begin{array}{cccc}
 -{\bf B^{\dagger}}  & 0 \\
 0 & {\bf I}
\end{array} \right) \right]
\left( \begin{array}{cccc} {\bm \Psi}_{i}\\
{\bm \Psi}_{i-1}
\end{array}
\right) = 0.
\end{equation}
Formally, the dimension of this problem is $2N_{\rm rs}$. There are a number of trivial
solutions, however, since ${\bf B},{\bf B^{\dagger}}$ are singular matrices. In appendix
\ref{reduc} it is shown how reduce the problem to its $2N \cdot W_y \cdot W_z$ non-trivial
solutions.

The non-trivial solutions of Eq.~(\ref{f7}) can be divided into two classes. The first class
comprises Bloch waves propagating to the right and evanescent waves decaying to the right;
the corresponding eigenvalues are denoted by $\lambda(+)$. The second class comprises Bloch
waves propagating to the left or evanescent waves decaying to the left; the eigenvalues are
denoted by $\lambda(-)$. The eigenvalues of the propagating waves have $|\lambda(\pm)|=1$ and
for evanescent waves $|\lambda(\pm)| \lessgtr 1$. The evanescent states come in pairs, since
it is easy to show that for every solution $\lambda(+)$ there is a corresponding solution
$\lambda(-)=1/\lambda^*(+)$. It can be shown that the propagating states also come in pairs,
i.e. for every right propagating wave $\lambda(+)$ there is a left propagating wave
$\lambda(-)$.\cite{pairs}

It makes sense to keep only those evanescent waves for which $1/\delta < |\lambda| < \delta$,
where $\delta$ is a sufficiently large number. States with $|\lambda|$ outside this interval
are extremely fast decaying or growing. Such states are not important in matching an ideal
wire to a scattering region. Typical of finite difference schemes there are also non-physical
solutions to Eq.~(\ref{f7}), which are related to so-called parasitic
modes.\cite{kreiss,branden} These are easily recognized and discarded since their
$|\lambda|$'s are either extremely small or large and thus fall outside the selected
interval. Moreover, these $|\lambda|$'s are very sensitive to the grid spacing and rapidly go
to 0 or $\infty$ if the grid spacing is decreased.

After filtering out the physical and useful solutions
Eq.~(\ref{f7}) we end up with $M$ pairs of solutions
$\lambda_m(\pm);m=1,\ldots,M$, where usually $M \ll N_{\rm rs}$.
We construct the normalized vectors ${\bf u}_m(\pm)$ from the
first $N_{\rm rs}$ elements of the eigenvectors of Eq.~(\ref{f7})
and form the $N_{\rm rs} \times M$ matrices
\begin{equation}\label{f03a}
{\bf U}(\pm)=({\bf u}_1(\pm) \cdots {\bf u}_{M}(\pm)).
\end{equation}
Choosing the cell $i=0$ as the origin, one then writes the general
solution ${\bm \Psi}_{0}$ in this cell as a linear combination of
these right- and left going modes
\begin{eqnarray}\label{f03}
{\bm \Psi}_{0}&=&{\bm \Psi}_{0}(+)+{\bm \Psi}_{0}(-) ,
\end{eqnarray}
where
\begin{eqnarray}\label{f03c}
{\bm \Psi}_{0}(\pm)&=&{\bf U}(\pm){\bf a}(\pm) =
\sum_{m=1}^{M}{\bf u}_m(\pm)a_m(\pm),
\end{eqnarray}
with ${\bf a}(\pm)$ vectors of arbitrary coefficients of dimension
$M$.

Defining the $M \times M$ diagonal eigenvalue matrices by
\begin{equation}\label{f03b}
({\bf \Lambda}(\pm))_{nm}=\delta_{nm}\lambda_m(\pm),
\end{equation}
and using the Bloch condition of Eq.~(\ref{f6}), the solution in
the other unit cells then can be expressed in a compact form
\begin{equation}\label{f04}
{\bm \Psi}_{i}={\bf U}(+)  {\bf \Lambda}^i(+){\bf a}(+) + {\bf
U}(-)  {\bf \Lambda}^i(-){\bf a}(-).
\end{equation}

In order to apply Ando's formalism,\cite{ando} it is advantageous
to slightly rewrite this. We define the $N_{\rm rs} \times N_{\rm
rs}$ matrices ${\bf F}(\pm)$ and $\widetilde{{\bf F}}(\pm)$ by
\begin{eqnarray}\label{f05}
{\bf F}(\pm){\bf U}(\pm) &=& {\bf U}(\pm){\bf \Lambda}(\pm),
\\
\label{f06} \widetilde{{\bf F}}(\pm){\bf U}(\pm) &=& {\bf
U}(\pm){\bf \Lambda}^{-1}(\pm).
\end{eqnarray}
Note that $\widetilde{{\bf F}}(\pm) \neq {\bf F}^{-1}(\pm)$ since
the $N_{\rm rs} \times M$ matrices ${\bf U}(\pm)$ are not square
(typically $M \ll N_{\rm rs}$). This presents no problem, however,
and explicit expressions for the matrices $\widetilde{{\bf
F}},{\bf F}$ are given in appendix~\ref{fmat}. They allow
Eq.~(\ref{f04}) to be rewritten in recursive form
\begin{eqnarray}\label{f07}
{\bm \Psi}_{i+1} &=& {\bf F}(+){\bm \Psi}_{i}(+) + {\bf F}(-){\bm \Psi}_{i}(-), \\
\label{f08} {\bm \Psi}_{i-1} &=& \widetilde{{\bf F}}(+){\bm
\Psi}_{i}(+) + \widetilde{{\bf F}}(-){\bm \Psi}_{i}(-),
\end{eqnarray}
either of which allows one to construct the full solution for the ideal wire, once the
boundary values are set, cf. Eq.~(\ref{f03}).

\subsection{Scattering problem}\label{scattering}
\begin{figure}[!tbp]
\includegraphics[width=8.0cm,keepaspectratio=true]{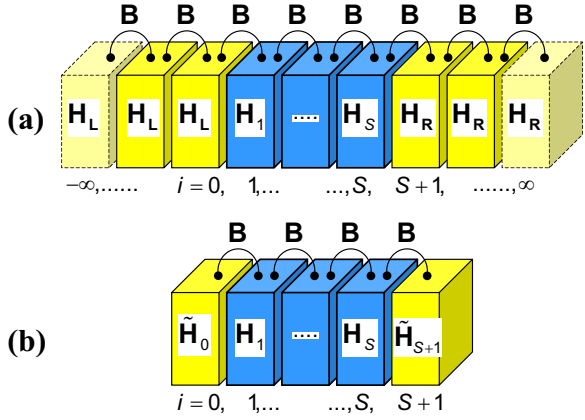}
\caption{(Color online) (a) Schematic representation of a quantum wire. The left (L) and
right (R) leads are ideal wires that span the cells $i=-\infty ,\ldots, 0$ and
$i=S+1,\ldots,\infty$, respectively. The scattering region spans cells $i=1,\dots,S$. (b) The
reduced problem spans the cells $i=0,\dots,S+1$, see Eq. (\ref{f016}).} \label{fig:struct}
\end{figure}

In a non-ideal quantum wire the potential is not periodic, which means that we have to solve
the Schr\"{o}dinger equation of Eq.~(\ref{f3}) with ${\bf H}_i$ depending upon the position
$i$ along the wire. The non-ideal region (the scattering region) is supposed to be finite,
spanning the cells $i=1,\dots,S$.\cite{cells} The left and right leads are ideal wires,
spanning the cells $i=-\infty ,\ldots, 0$ and $i=S+1,\ldots,\infty$, respectively. In the
ideal wires ${\bf H}_i$ does not depend on the position of the cell. However, the left lead
can be different from the right one, so we use the subscript L(R) to denote the
former(latter), i.e. ${\bf H}_{i} = {\bf H}_{\rm L},\; i<1$ and ${\bf H}_{i} = {\bf H}_{\rm
R},\; i>S$. A schematic picture of the structure is shown in Fig.~\ref{fig:struct}(a). We
solve Eq.~(\ref{f3}) over the whole space, $i=-\infty,\infty$, making use of the ideal wire
solutions of the previous section to reduce the problem to essentially the scattering region
only, see Fig.~\ref{fig:struct}(b).

For the solution in the left lead the recursion relation of
Eq.~(\ref{f08}) can be used. This gives for the cell $i=-1$
\begin{eqnarray}\label{f011}
{\bm \Psi}_{-1} &=& \widetilde{{\bf F}}_{\rm L}(+){\bm
\Psi}_{0}(+) +
\widetilde{{\bf F}}_{\rm L}(-){\bm \Psi}_{0}(-) \nonumber \\
&=& [\widetilde{{\bf F}}_{\rm L}(+)-\widetilde{{\bf F}}_{\rm
L}(-)]{\bm \Psi}_{0}(+) + \widetilde{{\bf F}}_{\rm L}(-){\bm
\Psi}_{0},
\end{eqnarray}
using Eq.~(\ref{f03}). The vector ${\bm \Psi}_{0}(+)$ describes a wave coming in from the
left. In a scattering problem this vector fixes the boundary condition. Eq.~(\ref{f011})
allows Eq.~(\ref{f3}) for $i=0$ to be written as
\begin{equation}\label{f012a}
( E{\bf I} - \widetilde{{\bf H}}_0)  {\bm \Psi}_{0} + {\bf
B^{\dagger}} {\bm \Psi}_{1} ={\bf Q}{\bm \Psi}_{0}(+),
\end{equation}
where
\begin{eqnarray}\label{f012b}
&&\widetilde{{\bf H}}_0 = {\bf H}_{\rm L} - {\bf B}\widetilde{{\bf
F}}_{\rm L}(-) \nonumber \\
&&{\bf Q}= {\bf B}[\widetilde{{\bf F}}_{\rm L}(-)-\widetilde{{\bf
F}}_{\rm L}(+)].
\end{eqnarray}

For the solution in the right lead we use the recursion relation
of Eq.~(\ref{f07}), which gives for the cell $i=S+2$
\begin{equation}\label{f013}
{\bm \Psi}_{S+2} = {\bf F}_{\rm R}(+){\bm \Psi}_{S+1}(+).
\end{equation}
Here we have assumed that in the right lead we have only a right going wave, which
corresponds to the transmitted wave. Eq.~(\ref{f013}) allows Eq.~(\ref{f3}) for $i=S+1$ to be
written as
\begin{equation}\label{f014}
(E{\bf I} - \widetilde{{\bf H}}_{S+1})  {\bm \Psi}_{S+1} + {\bf B}
{\bm \Psi}_{S} = 0,
\end{equation}
where
\begin{equation}\label{f014a}
\widetilde{{\bf H}}_{S+1} = {\bf H}_{\rm R} - {\bf
B^{\dagger}}{\bf F}_{\rm R}(+).
\end{equation}

Eqs.~(\ref{f012b}) and (\ref{f014a}) take care of the coupling of the scattering region to
the left and right leads. Eq.~(\ref{f3}) for $i=1,\ldots,S$ plus Eqs.~(\ref{f012a}) and
(\ref{f014}) form a complete set of equations from which the vectors ${\bm
\Psi}_{i};i=0,\ldots,S+1$ can be determined describing the waves in the scattering region.

The scattering reflection and transmission coefficients can be deduced from the amplitudes
immediately left and right of the scattering region, i.e. ${\bm \Psi}_{0}$ and ${\bm
\Psi}_{S+1}$. If we let the incoming wave consist of one specific mode,
${\bm\Psi}_{0}(+)={\bf u}_{{\rm L},n}$, i.e. $a_m(+)=\delta_{mn}$ in Eq.~(\ref{f03}), then
the generalized reflection and transmission probability amplitudes $r_{n'n}$ and $t_{n'n}$
are defined by
\begin{eqnarray}\label{f015}
{\bm \Psi}_{0}(-)=\sum_{n'=1}^{M_{\rm L}} {\bf u}_{{\rm L},n'}(-)  r_{n'n} \nonumber \\
{\bm \Psi}_{S+1}(+)=\sum_{n'=1}^{M_{\rm R}} {\bf u}_{{\rm
R},n'}(+) t_{n'n}.
\end{eqnarray}
Note that at this stage we include all evanescent and propagating modes since these form a
complete set to represent the states in the leads. We assume the lead states to be amplitude
normalized.

The reflection and transmission probability amplitudes $r_{n'n}$ and $t_{n'n}$ between all
possible modes form a $M_{\rm L} \times M_{\rm L}$ matrix ${\bf R}$ and a $M_{\rm R} \times
M_{\rm L}$ matrix ${\bf T}$, respectively. All elements of these matrices can be found in one
go by defining a $N_{\rm rs} \times M_{\rm L}$ matrix of all possible incoming modes, i.e.
\begin{equation}\label{f015a}
{\bf C}_{0}(+)={\bf U}_{\rm L}(+).
\end{equation}
Analogous to Eq.~(\ref{f015}) one then has
\begin{eqnarray}\label{f015b}
&&{\bf C}_{0}(-)={\bf C}_{0}-{\bf C}_{0}(+)={\bf U}_{\rm L}(-){\bf R} \nonumber \\
&&{\bf C}_{S+1}(+)={\bf C}_{S+1}={\bf U}_{\rm R}(+){\bf T}.
\end{eqnarray}
Eqs.~(\ref{f012a}), (\ref{f3}) and (\ref{f014}) then become
\begin{eqnarray}\label{f016}
\left\{ \begin{array}{ll}
 &  ( E{\bf I} - \widetilde{{\bf H}}_0){\bf C}_{0} + {\bf
B^{\dagger}} {\bf C}_{1} = {\bf Q}{\bf U}_{\rm L}(+) \\
 &  (E{\bf I} - {\bf H}_{i})  {\bf C}_{i} + {\bf B}{\bf C}_{i - 1}
+ {\bf B^{\dagger}} {\bf C}_{i + 1} = 0 \\
 &  (E{\bf I} - \widetilde{{\bf H}}_{S+1}){\bf C}_{S+1} +
{\bf B} {\bf C}_{S} = 0,
\end{array} \right.
\end{eqnarray}
$i=1,\ldots,S$. Solving this set of equations for ${\bf C}_{i};\,i=0,\ldots,S+1$ gives all
possible waves. From Eq.~(\ref{f015b}) one can then extract the generalized reflection and
transmission matrices ${\bf R}$ and ${\bf T}$. An efficient technique for solving the
equations is discussed in appendix \ref{scatt}.

In order to calculate the total transmission one has to select the transmission matrix
elements that refer to propagating modes and discard the ones that refer to evanescent modes.
This is easy, since the propagating modes have $|\lambda|=1$, see the discussion above
Eq.~(\ref{f03a}). The total transmission of Eq.~(\ref{f00}) is then given by
\begin{equation}\label{f017}
T(E) = \sum_{n=1,n'=1}^{m_{\rm L},m_{\rm R}} \frac{v_{{\rm
R},n'}}{v_{{\rm L},n}} \left|t_{n'n}\right|^2,
\end{equation}
where $v_{{\rm R},n'}$ and $v_{{\rm L},n}$ are the velocities in the $x$-direction of the
right propagating waves in the right and left lead in the modes $n'$ and $n$, respectively,
and $m_{\rm L},m_{\rm R}$ are the number of such modes. Introducing the velocities results
from flux normalizing the modes, which is required by current conservation.\cite{datta} The
velocities are given by the expression
\begin{equation}\label{f018}
v_n = -\frac{2a_x}{\hbar} {\rm Im} \left[ \lambda_n {\bf
u}_n^\dagger {\bf B}^\dagger {\bf u}_n \right] ,
\end{equation}
where subscripts L(R) need to be added for the left(right) leads. Eq.~(\ref{f018}) is derived
in appendix \ref{velocities}. The sign of the calculated velocities is used to distinguish
right from left propagating modes.

\subsection{Computational costs}\label{algorithm}
Our computational method can be summarized as follows. First, Eq.~(\ref{f7}) is solved in its
reduced form, Eq.~(\ref{a06}), to obtain the modes for both leads. The computing costs of
this step scale as $N_{\rm id}^3$, where $N_{\rm id}~=~\max(2N,L-N) \cdot W_y \cdot W_z$, see
appendix \ref{reduc}. These costs are small compared to the costs of solving the scattering
problem. The next step involves the selection of the physically relevant modes ${\bf u}_m$
and separate them into left $(+)$ and right $(-)$ going modes. The velocities are calculated
using Eq.~(\ref{f018}) and are used to distinguish left from right propagating states.
Evanescent states are classified as growing $(+)$ or decaying $(-)$ on account of their
eigenvalue. Subsequently, the ${\bf F}$-matrices are constructed, Eq.~(\ref{f05}), and the
matrix elements that define the boundary conditions on the scattering region are set, see
Eqs.~(\ref{f012b}) and (\ref{f014a}). The computing costs of these steps are minor.

The transmission matrix ${\bf T}$ is obtained by solving Eq.~(\ref{f016}) using the algorithm
of appendix \ref{scatt}. This is the most time consuming step. It scales as $S \cdot N_{\rm
rs}^3$, where $N_{\rm rs}=L \cdot W_y \cdot W_z$ is the number of grid points in the unit
cell and $S$ is the number of unit cells in the scattering region. Note that the scaling is
linear with respect to the size $S$ of the scattering region, which means that this algorithm
can be classified as $O(N)$. Finally, the total transmission and the conductance can be
obtained from Eqs.~(\ref{f017}) and (\ref{f00}).

\section{Results}\label{results}
\subsection{Numerical tests}\label{tests}

\begin{figure}[!tbp]
\includegraphics[width=8cm,keepaspectratio=true]{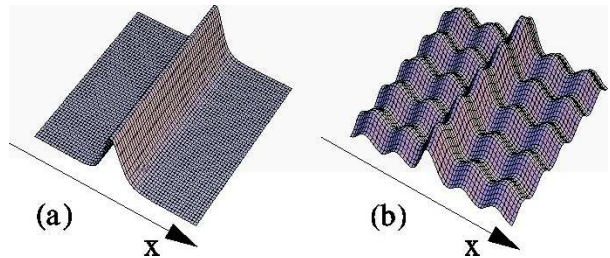}
\caption{(Color online) The potential of Eq.~(\protect\ref{r01}) in the $xy$-plane for the
cases (a) $V_0=0$ and (b) $V_0=V_1$.} \label{fig:cospot}
\end{figure}

In order to test the accuracy of our method we consider a system
described by the model potential
\begin{eqnarray}\label{r01}
V({\bf r}) &=& V_0 \left[ \cos(2\pi x/a)+\cos(2\pi y/a)+\cos(2\pi
z/a) \right] \nonumber \\
&+& \frac{V_1}{\cosh^2(\pi x/a)}.
\end{eqnarray}
The $V_0$ term describes an ideal wire by a simple three dimensional periodic potential with
periods $a_x=a_y=a_z=a$. The $V_1$ term describes a barrier in the propagation direction and
is a simple model for a tunnel junction. The potential is plotted in Fig.~\ref{fig:cospot}.
We solve the scattering problem for this system numerically in three dimensions by the method
outlined in Sec.~\ref{method}. Our results can be verified, however, since this potential is
in fact separable and limiting cases can be solved analytically. The solutions in the $y$-
and $z$-directions are Mathieu functions.\cite{abramowitz} If $V_1=0$ then the solutions in
the $x$-direction are also Mathieu functions. If $V_1 \neq 0$ but $V_0=0$ the scattering
problem can be solved analytically.\cite{landau} Finally, if $V_1 \neq 0$ and $V_0 \neq 0$
the solution in the $x$ direction can be obtained using the separability of the potential and
a standard numerical solver for the resulting ordinary differential equation in the
$x$-direction.\cite{mathematica} In the following the latter will be called the ``exact''
numerical solution.

\begin{figure}[!tbp]
\includegraphics[width=8.5cm,keepaspectratio=true]{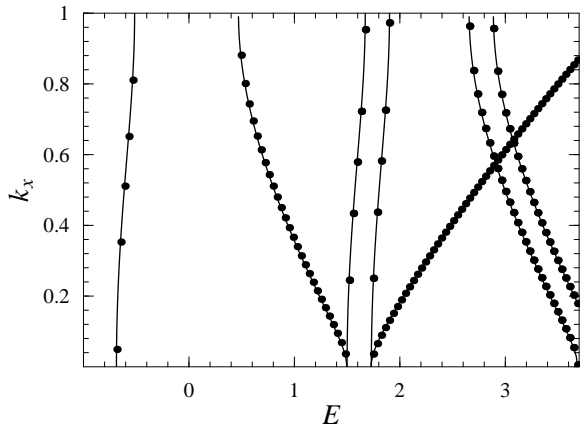}
\caption{The calculated wave number $k_x$ (in units of $\pi/a$) as function of the energy $E$
(in units of $V_0$) for an ideal wire. The points indicate the numerical results obtained
with $L,W_y,W_x=8$ and $N=4$. The solid line indicates the exact solution of the Mathieu
problem.} \label{fig:bands}
\end{figure}

As a first test we consider an ideal wire, i.e. $V_1=0$ in Eq.~(\ref{r01}). The potential is
separable and we can write the energy as
$E(k_x,k_y,k_z)=\epsilon_{n_x}(k_x)+\epsilon_{n_y}(k_y)+\epsilon_{n_z}(k_z)$, where
$\epsilon_n(k);\,k=-\pi/a,\ldots,\pi/a;\,n=0,1,\ldots$ are the eigenvalues of the Mathieu
problem.\cite{abramowitz} Fig.~\ref{fig:bands} shows part of the analytical band structure
for $(k_x,k_y,k_z)=(0,0,0)\rightarrow(\pi/a,0,0)$. It essentially consists of a superposition
of one-dimensional band structures $\epsilon_{n_x}(k_x)$ offset by energies
$\epsilon_{n_y}(0)+\epsilon_{n_z}(0);\,n_y,n_z=0,1,\ldots$.

The numerical band structure is obtained by solving Eq.~(\ref{f7}) in its reduced form,
Eq.~(\ref{a06}). To obtain the results shown in Fig.~\ref{fig:bands} we set ${\bf
k}_\parallel=(k_y,k_z)=(0,0)$, cf. Eq.~(\ref{a01b}), and determine the eigenvalues $\lambda$
in Eq.~(\ref{f7}) as a function of $E$. For the propagating states one can write
$\lambda=\exp(i k_x a)$. Plotting the calculated wave number $k_x$ as function of the energy
$E$ then allows to compare the results with the analytical band structure. The numerical
results shown in Fig.~\ref{fig:bands} are obtained using a grid of $L,W_y,W_z=8$ points per
period and a FDA with $N=4$. Although this grid is relatively coarse, we obtain a relative
accuracy on $k_x$ of $10^{-3}$.

This perhaps surprising accuracy is entirely due to the use of a high order FDA. To
illustrate this, Table~\ref{tab:conv1} shows the convergence of $k_x$ at a number of energies
$E$ as a function of the order $N$ of the FDA and the number of grid points $L$. These
particular results were obtained using the separability of the potential and solving the
problem numerically in the $x$-direction only, while using analytical solutions for the $y$-
and $z$-directions. For $N=6$ and $L=14$ the results are converged to within $10^{-7}$ of the
exact result. This is in sharp contrast to the results obtained with a simple first order
($N=1$) FDA, where a similar convergence can only be obtained at the cost of using two orders
of magnitude more grid points. Using such a large number of grid points in three dimensions
is entirely prohibitive because of the high computational costs involved. For example, aiming
at a moderate accuracy of $10^{-2}$, it is observed that for $N=4$ and $L=5$ the results are
markedly better than for $N=1$ and $L=14$. Yet in a three dimensional calculation, without
using the separability of the potential, the computing time required for the latter is two
orders of magnitude larger than for the former. It means that in order to solve a general
non-separable three dimensional problem with reasonable accuracy and computational costs, it
is vital to use a high order FDA.

\begin{table}[!]
\caption{$k_x(E)$ (in units of $\pi/a$) at values of $E$ (in units
of $V_0$) in the lowest two bands and in the first band gap of
Fig.~\ref{fig:bands}; in the band gap we find
$k_{x}=1+i\kappa_x$.} \label{tab:conv1}
\begin{tabular}{cccccc}
\toprule
$N$& $L$ &  $k_x(-0.6)$   & $k_x(1.0)$ & $\kappa_x(0.3)$  \\
\colrule
 1  &    7   & 0.694906   & 0.283091   & 0.283549   \\
    &   10   & 0.608958   & 0.322712   & 0.304238   \\
    &   14   & 0.571387   & 0.342011   & 0.312259   \\
    &  100   & 0.533341   & 0.359187   & 0.319658   \\
    & 1000   & 0.533084   & 0.359425   & 0.319688   \\
\colrule
 4  &  5     & 0.544347   & 0.355049   & 0.317777   \\
    &  7     & 0.533962   & 0.358927   & 0.319476   \\
    & 10     & 0.533149   & 0.359389   & 0.319673   \\
    & 14     & 0.533087   & 0.359425   & 0.319687   \\
\colrule
 6  & 7      & 0.533228   & 0.359337   & 0.319658   \\
    & 10     & 0.533086   & 0.359425   & 0.319688   \\
    & 14     & 0.533082   & 0.359428   & 0.319688   \\
\colrule
 {\rm Exact} & & 0.533082 & 0.359428   & 0.319688   \\
\botrule
\end{tabular}
\end{table}

\begin{figure}[!tbp]
\includegraphics[width=8.5cm,keepaspectratio=true]{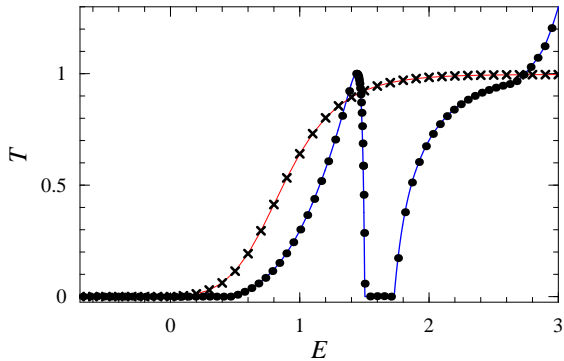}
\caption{(Color online) The total transmission as function of energy (in units of $V_1$) for
the two cases $V_0=0$ (crosses) and $V_0=V_1$ (dots). In both cases ${\bf k}_{\|}=(0,0)$. The
solid lines represent the analytical solution for $V_0=0$ and the ``exact'' numerical
solution for $V_0=V_1$.} \label{fig:trans1}
\end{figure}

Next we consider the scattering problem and calculate the total transmission for the case
where $V_1 \neq 0$. The size of the scattering region is set to $Sa$ and outside this region
the scattering potential (the last term of Eq.~(\ref{r01})) is set to zero. With $S=6$ the
results are extremely well converged. As an example we have calculated the transmission at
normal incidence, i.e. ${\bf k}_\parallel=(k_y,k_z)=(0,0)$. The crosses marked $V_0=0$ in
Fig.~\ref{fig:trans1} represent the numerical results for the transmission of the
corresponding potential, obtained with a $L,W_y,W_z=8$ grid and a $N=4$ FDA. This scattering
problem can also be solved analytically, \cite{landau} and the analytical and numerical
transmission probabilities agree within $10^{-4}$.

Fig.~\ref{fig:trans1} also shows the transmission for the case where $V_0=V_1$ as calculated
numerically using the same parameters as before, i.e. $S=6;L,W_y,W_z=8;N=4$. This scattering
problem can only be solved semi-analytically; Mathieu solutions are used in $y$- and
$z$-directions, and the (ordinary) differential equation for the $x$-direction is solved
``exactly'' using an accurate standard numerical solver.\cite{mathematica} Again the
``exact'' and numerical transmission probabilities agree within $10^{-4}$. Compared to the
$V_0=0$ case it is observed that the influence of the periodic potential of the leads upon
the transmission is large. For $V_0=V_1$ the electronic states in the leads are far from free
electron-like, see Fig.~\ref{fig:bands}. In particular, the transmission drops to zero if the
energy is inside a band gap, because there are no lead states of that energy.

\begin{figure}[!tbp]
\includegraphics[width=8.5cm,keepaspectratio=true]{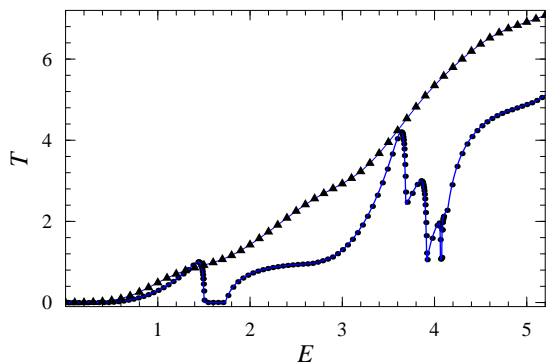}
\caption{(Color online) The total transmission as function of energy (in units of $V_1$) for
the two cases ${\bf k}_{\|}=(0,0)$ (dots) and ${\bf k}_{\|}=(0.47,0.21)\pi/a$ (triangles). In
both cases $V_0=V_1$. The solid lines represent the ``exact'' numerical solution.}
\label{fig:trans2}
\end{figure}

The numerical calculations accurately capture the transmission curve over a large energy
range, as is shown in Fig.~\ref{fig:trans2}. The transmission generally increases with energy
due to the increasing number of channels, see Fig.~\ref{fig:bands}. Since the density of
states peaks at the band edges, the transmission peaks at the corresponding energies. The
transmission depends very much  upon ${\bf k}_{\|}$ as can be observed in
Fig.~\ref{fig:trans2}, where the transmission for normal incidence, ${\bf k}_{\|}=(0,0)$, can
be compared to that for ${\bf k}_{\|}=(0.47,0.21)\pi/a$ (an arbitrary point in the Brillouin
zone). The difference between the two curves can be easily understood from the band structure
of the leads. In particular, for ${\bf k}_{\|}=(0.47,0.21)\pi/a$ there are no band gaps for
$E>0.14V_0$.

\begin{figure}[!tbp]
\includegraphics[width=7.5cm,keepaspectratio=true]{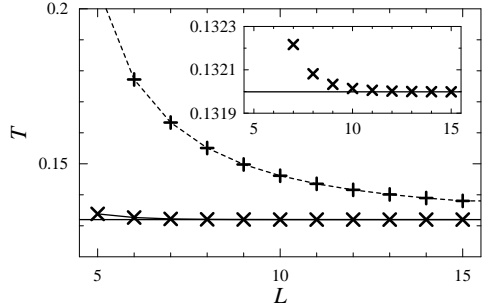}
\caption{The total transmission $T$ as function of the grid size $L=W_y=W_z$ for a simple
$N=1$ FDA,`$+$' top curve, and for a $N=4$ FDA, `$\times$' bottom curve. The horizontal line
represents the ``exact'' value $T=0.132$. The inset shows the $N=4$ curve on a finer scale
for $T$.} \label{fig:conv}
\end{figure}

To demonstrate the convergence of the numerical calculations, Fig.~\ref{fig:conv} shows the
total transmission as function of the sampling density  $L=W_y=W_z$ for a simple $N=1$ and a
high order $N=4$ FDA. The results shown are for one particular ${\bf
k}_\parallel=(k_y,k_z)=(0.47,0.21)\pi/a$ and energy $E=0.895$ $V_0$, but the convergence at
other ${\bf k}_\parallel$-points and energies is very similar. The number of propagating
channels at this ${\bf k}_\parallel$-point and energy is two, but the total transmission is
only $T=0.132$, which means that the barrier is largely reflecting. We conclude that the
accuracy of the three-dimensional calculation depends very strongly upon the order of the
FDA. For $N=1,L=15$, the transmission is converged on a scale of $10^{-2}$ only, but for
$N=4$ it is converged on a scale of $10^{-3}$ already for $L=8$, see the inset of
Fig.~\ref{fig:conv}. A high order FDA thus enables the use of a much coarser real space grid.
Since the computational costs scale with the number of real space grid points $N_{\rm
rs}=L^3$ as $N_{\rm rs}^3=L^9$, this demonstrates the strength of using a high order FDA.

\subsection{Sodium atomic wires}\label{realstuff}
\begin{figure}[!tbp]
\includegraphics[width=8.5cm,keepaspectratio=true]{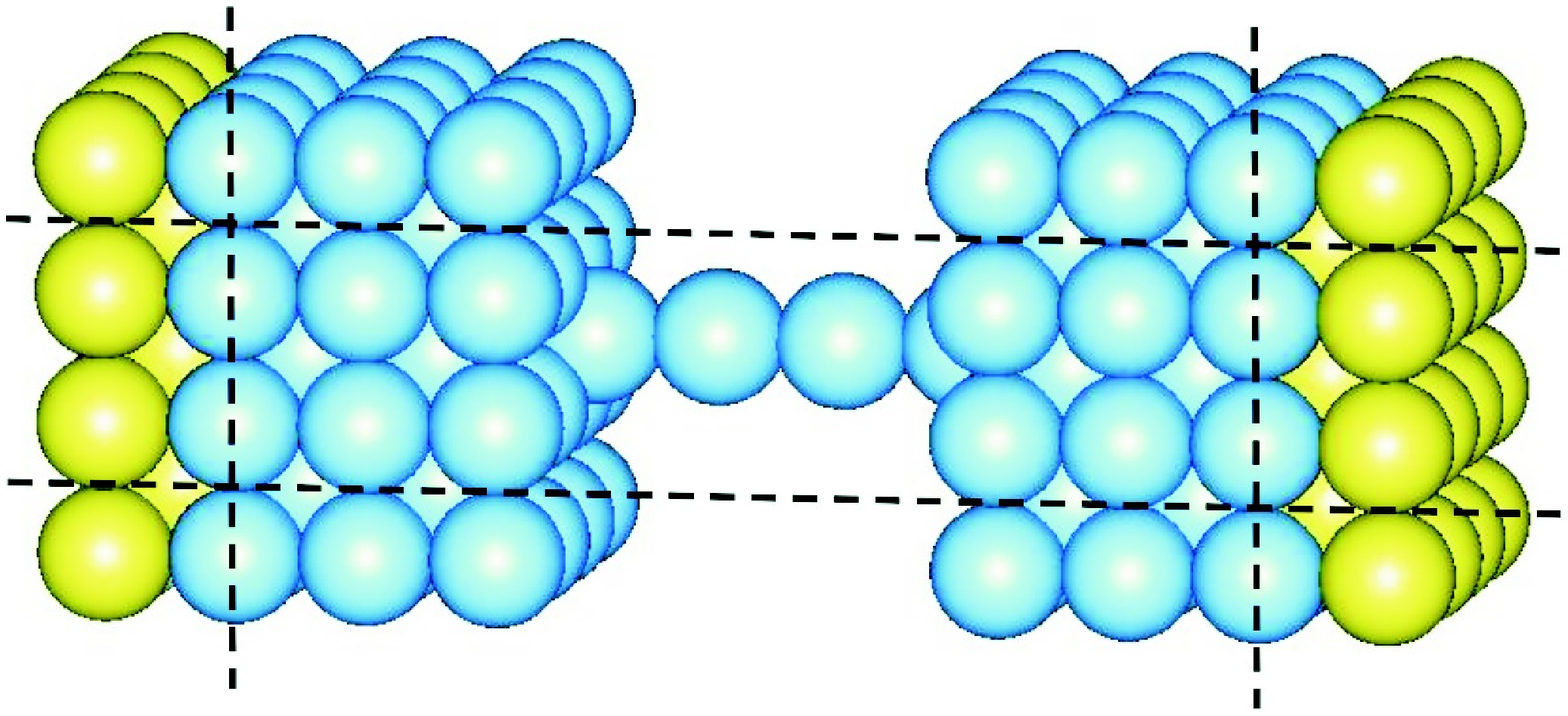}
\caption{(Color online) Structure of an atomic wire consisting of 4 sodium atoms between two
sodium leads terminated by (100) surfaces. The scattering region is bounded by the vertical
lines and the lateral supercell by the horizontal lines. Bulk atoms are indicated by yellow
(light grey) balls and atoms in the scattering region by blue (dark grey) balls,
respectively.} \label{fig:nawire}
\end{figure}

We have calculated the electronic transport in sodium atomic wires as examples of more
complex systems. Our model of a sodium wire consists of left and right leads composed of bulk
(bcc) sodium metal terminated by a (100) surface, connected by a straight wire of sodium
atoms, as is shown in Fig.~\ref{fig:nawire}. The atoms in the leads are positioned according
to the bcc structure of bulk sodium, with the cell parameter fixed at the experimental value
of 7.984 $a_0$.\cite{nalatt} The atoms in the wire are fixed at their (bulk) nearest neighbor
distance of 6.915 $a_0$. Since geometry relaxation at the Na(100) surface is very
small,\cite{quong} and calculations using jellium electrodes have shown that the conductance
of a sodium wire is not very sensitive to its geometry,\cite{kobayashi} we have refrained
from optimizing the geometry. Perpendicular to the wire we apply periodic boundary conditions
using a $2 \times 2$ lateral supercell, which has a lattice parameter of 15.968 $a_0$.

If DFT is used to model the electronic structure, Eq. (\ref{f0}) corresponds to the Kohn-Sham
equation. The one-electron potential $V(\mathbf{r})$ in this equation is then given by the
sum of the nuclear Coulomb potentials or pseudopotentials, and the electronic Hartree and
exchange-correlation potentials. The latter two depend upon the electronic charge density. In
linear response the charge density remains that of the ground state, allowing the electronic
potentials to be obtained from a self-consistent ground state calculation. In these
calculations we employ supercells containing a slab of 13 layers to represent the bulk and
surface of the leads, and a wire of $n$ atoms, see Fig.~\ref{fig:nawire}. We use the local
density approximation,\cite{perdew} and represent the ion cores of the sodium atoms by a
local pseudopotential.~\cite{TM} The valence electronic wave functions are expanded in a
plane wave basis set with a kinetic energy cutoff of 16 Ry. The lateral Brillouin zone is
sampled with a $8 \times 8$ ${\bf k}_{\|}$-point grid, using a temperature broadening with
$kT_{el}=0.1$ eV.\cite{neugebauer}

\begin{figure}[!tbp]
\includegraphics[width=8.5cm,keepaspectratio=true]{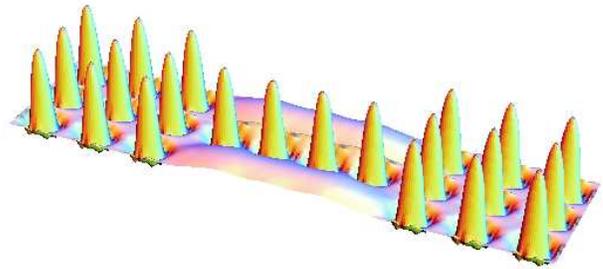}
\caption{(Color online) The effective potential for valence electrons in the $xy$-plane of
the sodium atomic wire shown in Fig~\protect\ref{fig:nawire}. Most prominent are the strongly
repulsive core regions and the attractive valence regions of the atoms. The difference
between the maximum and minimum values of this potential is $32.4$ eV. The Fermi level is at
$8.5$ eV above the potential minimum.} \label{fig:napotential}
\end{figure}

We want to calculate the conductance for various lengths of the atomic wire, so for each
length $n$ we perform a self-consistent supercell calculation to generate the one-electron
potential. By expressing the latter in a plane wave basis, Fourier interpolation can be used
to obtain a representation on any real space grid required for the transport calculations,
cf. Eq. (\ref{f02}). An example of the effective potential for valence electrons of a sodium
atomic wire is shown in Fig.~\ref{fig:napotential}.

For the transport calculations we use a scattering region comprising the atomic wire and the
surface regions of the left and right leads. Both surface regions consist of 5 atomic layers,
see Fig.~\ref{fig:nawire}. Periodic boundary conditions and a $2 \times 2$ lateral supercell
perpendicular to the wire are applied. The potential in the scattering region is extracted
from the slab calculations. Outside the scattering region we assume that the leads consist of
bulk sodium. The potential for the leads and the value of the Fermi energy are extracted from
a bulk calculation. The average bulk potential is lined up with the average potential in the
middle of the supercell slab.\cite{fall} We have checked that the spatial dependence of the
bulk potential is virtually identical to that of the potential in the middle of the slab.
This means that the connection between the leads and the scattering region is smooth. There
are no discontinuities in the potential in the boundary regions that could cause spurious
reflections.

\begin{figure}[!tbp]
\includegraphics[width=8.5cm,keepaspectratio=true]{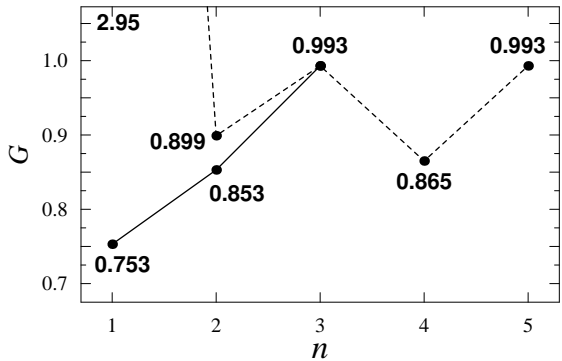}
\caption{Top curve: conductance $G(E_F)$ at the Fermi level (in units of $e^2/\pi \hbar$) of
a sodium atomic wire as a function of the number of atoms $n$ in the wire. Bottom curve:
conductance of a sodium atomic wire relative to vacuum tunneling conductance between two
electrodes without wire.} \label{fig:naconductance}
\end{figure}

Fig.~\ref{fig:naconductance} shows the calculated conductance $G(E_F)$ at the Fermi level as
a function of the length $n$ of the sodium atomic wire. The conductance is calculated using a
grid spacing along $x,y$ and $z$ directions of $h_{x,y,z}=0.67\,a_{0}$ (giving
$L=6,W_y=W_z=24$) and a FDA of order $N=4$.\cite{dimremark} The $2\times 2$ lateral Brillouin
zone is sampled by a $12 \times 12$ uniform ${\bf k}_{\|}$-point grid. With these parameters
the calculated conductances have converged to well within $10^{-3}$ $G_0$ (where
$G_0=e^2/(\pi\hbar)$).\cite{convremark} All wires have a conductance close to unity, except
the one-atom ($n=1$) wire. The high conductance for the one-atom wire is foremost due to
tunneling between the left and right leads through vacuum. The latter can be calculated by
omitting the wire and otherwise keeping the geometry fixed. Tunneling through vacuum leads to
a conductance of 2.20 $G_0$ per $2\times 2$ surface cell. Vacuum tunneling decreases fast
with the distance between left and right lead surfaces. At a distance corresponding to the
two-atom wire it gives a conductance of 0.047 $G_0$; at distances corresponding to longer
wires this conductance is negligible. Subtracting these vacuum tunneling values gives the
lower curve in Fig.~\ref{fig:naconductance}.

With the exception of the one-atomic wire, the conductances are close to unity. This is
perhaps not surprising, since within a tight-binding model atomic wires consisting of a
monovalent atom like sodium are expected to have one open channel. For perfectly transmitting
contacts this would give a conductance of 1 $G_0$.\cite{levy} Our calculated conductances for
$n>1$ are less than 15\% smaller than this value, demonstrating that this transmission is
indeed very high. The conductance of the one-atomic wire, relative to the vacuum tunneling
conductance at this distance, is significantly lower than unity. The electronic structure of
a single atom between two electrodes is substantially distorted from the simple single open
channel model.\cite{lang}

On a finer scale we find evidence of an even-odd oscillation in the conductance obtained in
previous studies.\cite{lang,sim,havu,tsukamoto,stockbroe} The conductance for wires with $n$
even tends to be lower than for those with $n$ odd. In simple tight-binding terms
odd-numbered atomic wires have a non-bonding level that tends to line up with the Fermi level
of the leads, which gives a high transmission. For an even-numbered atomic wire on the other
hand the Fermi level tends to fall in the gap between the bonding and anti-bonding levels of
the wire, resulting in a lower transmission.\cite{jacobsen} The size of the even-odd
oscillation in the conductance depends of course upon the nature of the contacts between the
atomic wire and the leads. Good contacts broaden the levels of the atomic wire into wide
resonances, which tends to suppress the even-odd oscillation. Our wires have good contacts,
but the even-odd oscillation remains distinctly visible.\cite{potremark}

\section{Summary}\label{summary}
We have formulated and implemented a new numerical technique for calculating electronic
transport in quantum wires and tunnel junctions in the linear response regime, starting from
Landauer's scattering formalism. It is based upon a real space grid representation of the
scattering problem. Dividing space into left and right ideal leads and a scattering region,
the problem is solved by \textit{wave function matching} (WFM). First all propagating and
evanescent Bloch modes of the leads are calculated. Subsequently the states in the scattering
region are forced to match to the Bloch modes of the leads. This directly leads to the
transmission matrix, which contains the transmission probability amplitudes between all modes
of the left and right leads, and to the conductance. The computing costs of this algorithm
scale linearly with the size of the scattering region.

It is shown that the use of a high order finite difference approximation for the kinetic
energy operator leads to a high accuracy and efficiency. This is demonstrated for a model
potential by benchmarking the technique against analytical and numerically ``exact''
solutions. The method is then applied to calculate the conductance in sodium atomic wires,
where the potential in the wire and in the bulk sodium leads is obtained from self-consistent
DFT calculations.

\acknowledgments

The authors acknowledge helpful discussions with P. J. Kelly and M. Zwierzycki. The work of
PAK is financially supported by a grant from the `Nederlandse Organisatie voor
Wetenschappelijk Onderzoek (NWO), gebied Chemische Wetenschappen (CW)'. The work of GB is
part of the research program of the `Stichting voor Fundamenteel Onderzoek der Materie
(FOM)', financially supported by NWO. The `Stichting Nationale Computerfaciliteiten (NCF)' is
acknowledged for granting computer time.

\appendix
\section{Computational Details}\label{details}
\subsection{H and B matrices}\label{handb}
In this section the matrices ${\bf H}_i$ and ${\bf B}$, introduced in Sec.~\ref{finite}, are
presented explicitly. In order not to complicate the notation the subscript $i$ is dropped;
all quantities refer to a single cell. For a wire with a finite cross-section in the $yz$
plane, the matrix ${\bf H}$ is real and symmetric and has the form
\begin{equation}\label{a01}
{\bf H} = \left( \begin{array}{cccccccc}
  {\bf h}_{1} & -{\bm \beta}_{1} & \ldots & -{\bm \beta}_{N}& 0 & \ldots & 0 & 0 \\
  -{\bm \beta}_{1} & {\bf h}_{2} & \ldots & -{\bm \beta}_{N-1}& -{\bm \beta}_{N} & \ldots & 0 & 0 \\
  \vdots & \vdots & \vdots & \vdots & \vdots & \vdots  & \vdots & \vdots  \\
  0 & 0 & \ldots & 0 & 0 &\ldots & {\bf h}_{L - 1} & -{\bm \beta}_{1} \\
  0 & 0 & \ldots & 0 & 0 &\ldots & -{\bm \beta}_{1} & {\bf h}_{L}
\end{array} \right) .
\end{equation}
Here $N$ is the order of the finite difference formula used, see Eq.~(\ref{f01}). We assume
that the $x$-axis is in the direction of the wire. $L$ is the number of grid points in the
$x$-direction of the unit cell defined by the periodic potential.

The submatrices ${\bf h}_{n}$ and ${\bm \beta}_{n}$ are of dimension $W_y \times W_z$, which
is the number of grid points in the cross section of the wire. Denoting
$(k,l)=k+(l-1)W_y;k=1,\ldots,W_y;l=1,\ldots,W_z$ as the compound index covering the grid
points in the cross section, the non-zero elements of these matrices are easily derived from
Eq.~(\ref{f02}).
\begin{eqnarray}\label{a01a}
({\bf h}_j)_{(k,l),(k,l)}&=&V_{j,k,l}-\left(t^{x}_{0}+t^{y}_{0}+t^{z}_{0}\right) \nonumber \\
({\bf h}_j)_{(k,l),(k+n,l)}&=&-t^{y}_{n},\; n \neq 0 \nonumber \\
({\bf h}_j)_{(k,l),(k,l+n')}&=&-t^{z}_{n'},\; n' \neq 0 \nonumber \\
({\bm \beta}_j)_{(k,l),(k,l)}&=& t^{x}_{j},
\end{eqnarray}
where $ -N\leq n,n' \leq N$ and $1 \leq k,k+n \leq W_y;1\leq l,l+n' \leq W_z$;$1\leq j\leq
N$. Note that in writing down these matrices we have assumed that $N<L,W_y,W_z$. In practical
calculations on realistic systems this will always be the case.

The matrix ${\bf B}$ has the same dimension as ${\bf H}$, but it
is upper triangular
\begin{eqnarray}\label{a02}
{\bf B} = \left( \begin{array}{ccccccc}
 0 & \ldots & 0 & {\bm \beta}_{N} & {\bm \beta}_{N-1} & \ldots & {\bm \beta}_{1} \\
 0 & \ldots & 0 & 0     & {\bm \beta}_{N}   & \ldots & {\bm \beta}_{2} \\
 \vdots & \vdots & \vdots  & \vdots & \vdots & \vdots & \vdots \\
 0 & \ldots & 0 & 0     & 0       & \ldots & {\bm \beta}_{N} \\
 \vdots & \vdots & \vdots  & \vdots & \vdots & \vdots & \vdots \\
 0 & \ldots & 0 & 0     & 0       & \ldots & 0
 \end{array} \right) .
\end{eqnarray}

For a wire that is periodic in the $yz$ plane, the wave functions in Eq.~(\ref{f02}) must
obey Bloch conditions. That is, $\Psi_{j,k+W_y,l}=e^{ik_ya_y}\Psi_{j,k,l}$ and
$\Psi_{j,k,l+W_z}=e^{ik_za_z}\Psi_{j,k,l}$, where $a_y,a_z$ are the periods in the $y$ and
$z$ directions, and $(k_y,k_z)={\bf k}_\parallel$ is the Bloch wave vector in the $yz$ plane.
These Bloch conditions in the $yz$ plane can be taken into account by defining the blocks
\begin{eqnarray}\label{a01b}
({\bf h'}_j)_{(k,l),(k+W_y+n,l)}&=&-t^{y}_{n}
e^{-i k_ya_y},\; n=-N,\ldots,-k \nonumber \\
({\bf h'}_j)_{(k,l),(k-W_y+n,l)}&=&-t^{y}_{n}
e^{i k_ya_y},\; n=W_y-k,\ldots,N \nonumber \\
({\bf h'}_j)_{(k,l),(k,l+W_z+n')}&=&-t^{z}_{n'}
e^{-i k_za_z},\; n'=-N,\ldots,-l \nonumber \\
({\bf h'}_j)_{(k,l),(k,l-W_z+n')}&=&-t^{z}_{n'} e^{i k_za_z},\;
n'=W_z-l,\ldots,N. \nonumber \\
\end{eqnarray}
The matrix ${\bf H}({\bf k}_\parallel)$, which is obtained by substituting ${\bf h}_j$ by
${\bf h}_j+{\bf h}'_j;j=1,\dots,L$ in Eq.~(\ref{a01}) describes a wire that is periodic in
the $yz$ plane with solutions corresponding to a Bloch vector ${\bf k}_\parallel$. This
matrix is (complex) Hermitian.

\subsection{Ideal wire}\label{reduc}
For an ideal wire, which has a periodic potential along the wire, Eq.~(\ref{f7}) has to be
solved to find the propagating and the evanescent waves. The precise form of the submatrices
in Eqs.~(\ref{a01}) and (\ref{a02}) is not important in the following discussion. For ease of
notation we only mention the dimensions $L$ (the number of grid points in the $x$-direction)
and $N$ (the order of the finite difference expression) explicitly and treat the wire as
quasi one-dimensional. To find the dimensions of the matrices in the three-dimensional case,
one simply has to multiply the dimensions mentioned below by $W_y \times W_z$.

Eq.~(\ref{f7}) is a generalized eigenvalue problem of dimension $2L$. Because the matrix
${\bf B}$ is singular it has a number of trivial solutions $\lambda=0$ and $\lambda=\infty$.
By using a partitioning technique we will eliminate these trivial solutions and reduce the
problem to the $2N$ non-trivial solutions. The key point is to split the vectors ${\bm
\Psi}_{i}$ into two parts containing the first $L-N$ and last $N$ elements, respectively. The
two parts are denoted by the subscripts 1 and 2. Splitting the matrices ${\bf H}$ and ${\bf
B}$ in the same way one gets
\begin{equation}\label{a03}
{\bm \Psi}_{i} = \left( \begin{array}{c}
{\bm \Psi}_{i,1} \\
{\bm \Psi}_{i,2}
\end{array} \right) \,;\;
{\bf H} = \left( \begin{array}{cc}
 {\bf H}_{11} & {\bf H}_{12}  \\
 {\bf H}_{21} & {\bf H}_{22}
\end{array} \right) \,;\;
{\bf B} = \left( \begin{array}{cc}
 0            & {\bf B}_{12}  \\
 0            & {\bf B}_{22}
\end{array} \right) .
\end{equation}
Note the special form of the matrix ${\bf B}$.

This splitting allows Eq.~(\ref{f7}) to be written in the form
\begin{eqnarray}\label{a04}
 &&\left( \begin{array}{cccc}
 E{\bf I}_{11} - {\bf H}_{11} & - {\bf H}_{12} & 0        & {\bf B}_{12} \\
- {\bf H}_{21} + \lambda {\bf B}^{\dagger}_{21} & E{\bf I}_{22} -
{\bf H}_{22} +
\lambda {\bf B}^{\dagger}_{22} & 0                          & {\bf B}_{22} \\
 {\bf I}_{11} & 0            & - \lambda {\bf I}_{11}     & 0            \\
 0            & {\bf I}_{22} & 0                          & - \lambda {\bf I}_{22}
\end{array} \right) \times \nonumber \\
&& \left( \begin{array}{c}
{\bm \Psi}_{i,1} \\
{\bm \Psi}_{i,2} \\
{\bm \Psi}_{i-1,1} \\
{\bm \Psi}_{i-1,2}
\end{array}
\right) = 0,
\end{eqnarray}
From this expression it is clear that the component ${\bm \Psi}_{i-1,1}$ only enters the
problem in a trivial way as ${\bm \Psi}_{i-1,1}=1/\lambda \times {\bm \Psi}_{i,1}$. It can be
eliminated by deleting the third row and column in Eq.~(\ref{a04}).

Furthermore the first row of the matrix does not depend upon the eigenvalue $\lambda$.
Writing out the multiplication for the first row explicitly, one finds an expression for
${\bm \Psi}_{i,1}$
\begin{equation}\label{a05}
{\bm \Psi}_{i,1} = \left( E{\bf I}_{11}-{\bf H}_{11}\right)^{-1}
 \left({\bf H}_{12}  {\bm \Psi}_{i,2} - {\bf B}_{12}  {\bm
\Psi}_{i-1,2} \right) .
\end{equation}

This can be used to eliminate ${\bm \Psi}_{i,1}$ from
Eq.~(\ref{a04}) to arrive at the equation
\begin{equation}\label{a06}
\left[ \left( \begin{array}{cc}
 {\bf A}_{11}      & {\bf A}_{12} \\
 {\bf I}_{22}      & 0
\end{array} \right) \right.
- \left. \lambda \left( \begin{array}{cc}
 {\bf S}_{11} & {\bf S}_{12} \\
 0            & {\bf I}_{22}
\end{array} \right) \right]
\left( \begin{array}{c}
{\bm \Psi}_{i,2} \\
{\bm \Psi}_{i-1,2}
\end{array}
\right) = 0,
\end{equation}
with
\begin{eqnarray}\label{a07}
&& {\bf A}_{11} = E{\bf I}_{22}-{\bf H}_{22} - {\bf H}_{21} \left(
E{\bf I}_{11}-{\bf
H}_{11}\right)^{-1}  {\bf H}_{12}\nonumber \\
&& {\bf A}_{12} = -{\bf H}_{21}  \left( E{\bf
I}_{11}-{\bf H}_{11}\right)^{-1}  {\bf B}_{12}  \nonumber \\
&& {\bf S}_{11} = -  {\bf B}^{\dagger}_{22} - {\bf
B}^{\dagger}_{21} \left( E{\bf I}_{11}-{\bf H}_{11}\right)^{-1}
{\bf H}_{12}  \nonumber \\
&& {\bf S}_{12} = - {\bf B}^{\dagger}_{21} \left( {E{\bf
I}_{11}-\bf H}_{11}\right)^{-1}  {\bf B}_{12} .
\end{eqnarray}

Eq.~(\ref{a06}) is a generalized eigenvalue problem of dimension $2N$ that can be solved
using standard numerical techniques.\cite{golub} In general it gives $2N$ eigenvalues
$\lambda_m$ and eigenvectors ${\bf u}_m$. As mentioned in the text, some of these solutions
are non-physical,\cite{kreiss,branden} others represent extremely fast growing or decaying
waves. Both of these classes of unwanted solutions are easily filtered out by demanding that
$1/\delta < |\lambda| < \delta$, where $\delta$ is some threshold value. We use this
criterion to select the physically relevant solutions, which are then separated into $M$
right-going and $M$ left-going solutions. These are used to construct the matrices of
Eqs.~(\ref{f03a}) and (\ref{f03b}) which contain all the information required to describe the
ideal wire.

The computational cost of solving Eq.~(\ref{a06}) scales as $(2N)^3$, whereas the cost of
computing the matrices of Eq.~(\ref{a07}) basically scales as $(L-N)^3$ (which is the cost of
the matrix inversion involved). Depending on the relative sizes of $L$ and $N$ one of these
two steps is dominant.

\subsection{F matrices}\label{fmat}
In this section explicit expression for the matrices ${\bf F}$ and $\widetilde{{\bf F}}$ are
given, see Eqs.~(\ref{f05}) and (\ref{f06}). Following Eq.~(\ref{f03a}), we denote the
propagating and evanescent modes of the ideal wire by ${\bf u}_m;m=1,\ldots,M$, where $M <
N_{\rm rs}$ and $N_{\rm rs}$ is the dimension of the vectors. For clarity of notation we omit
the labels $\pm$ for right- and left-going modes here. As in Eq.~(\ref{f03a}) we form the
$N_{\rm rs} \times M$ matrix
\begin{eqnarray}\label{b01}
{\bf U}&=&({\bf u}_1 \cdots {\bf u}_M) \\
&=& \left( \begin{array}{ccc}
 u_{11} & \ldots & u_{1M} \\
 \vdots &        & \vdots \\
 u_{N_{\rm rs}1} & \ldots & u_{N_{\rm rs}M}
\end{array} \right) .
\end{eqnarray}
The mode vectors ${\bf u}_m$ are in general non-orthogonal and we can form the $M \times M$
(positive definite) overlap matrix with elements
\begin{equation}\label{b02}
S_{mn}={\bf u}_m^{\dagger}{\bf u}_n \equiv \langle {\bf u}_m |
{\bf u}_n \rangle .
\end{equation}
This allows us to construct the dual basis $\widetilde{{\bf u}}_m;m=1,\ldots,M$
\begin{equation}\label{b03}
\widetilde{{\bf u}}_m = \sum_{n=1}^M S^{-1}_{mn}{\bf u}_n,
\end{equation}
with properties
\begin{equation}\label{b03a}
\langle \widetilde{{\bf u}}_m | {\bf u}_n \rangle = \langle {\bf
u}_m | \widetilde{{\bf u}}_n \rangle = \delta_{mn} .
\end{equation}
Now define the $M \times N_{\rm rs}$ matrix
\begin{eqnarray}\label{b04}
\widetilde{{\bf U}} &=& (\widetilde{{\bf u}}_1 \cdots \widetilde{{\bf u}}_M)^{\dagger} \nonumber \\
&=& \left(
\begin{array}{ccc}
 \widetilde{u}_{11}^* & \ldots & \widetilde{u}_{N_{\rm rs}1}^* \\
 \vdots             &        & \vdots \\
 \widetilde{u}_{1M}^* & \ldots & \widetilde{u}_{N_{\rm rs}M}^*
\end{array} \right) .
\end{eqnarray}
$\widetilde{{\bf U}}$ is called the pseudo-inverse of ${\bf U}$; note that $\widetilde{{\bf
U}} {\bf U} = {\bf I}_M$, where ${\bf I}_M$ is the $M \times M$ identity matrix.\cite{golub}

Defining the matrix
\begin{equation}\label{b05}
{\bf F} = {\bf U}{\bf \Lambda}\widetilde{{\bf U}} ,
\end{equation}
it is easy to show that it is a solution to Eq.~(\ref{f05}). ${\bf
F}$ is in fact a matrix that projects onto the space spanned by
the modes, as is easily demonstrated by writing Eq.~(\ref{b05}) as
\begin{equation}\label{b06}
{\bf F} = \sum_{m=1}^M |{\bf u}_{m} \rangle  \lambda_{m} \langle
\widetilde{{\bf u}}_m| ,
\end{equation}
making use of Eqs.~(\ref{f03b}) and (\ref{b01})-(\ref{b04}). In a
similar way a solution to Eq.~(\ref{f06}) is formed by
\begin{eqnarray}\label{b07}
\widetilde{{\bf F}} &=& {\bf U}{\bf \Lambda}^{-1}\widetilde{{\bf
U}} \nonumber \\
&=& \sum_{m=1}^M |{\bf u}_{m} \rangle  \lambda_{m}^{-1} \langle
\widetilde{{\bf u}}_m| .
\end{eqnarray}
Note that $\widetilde{{\bf F}}={\bf F}^{-1}$ only if $M=N_{\rm
rs}$, but since $M \leq N \times W_x \times W_y < N_{\rm rs}$ (see
the previous section) this will never be the case.

\subsection{Scattering problem}\label{scatt}
The scattering problem is described by Eq.~(\ref{f016}). It is conveniently written in matrix
form as
\begin{eqnarray}\label{c01}
\left( \begin{array}{ccccc} {\bf A}_{0} & {\bf B^{\dagger}} & 0 &
\ldots
& 0 \\
{\bf B} & {\bf A}_{1} & {\bf B^{\dagger}} & \ldots & 0 \\
0 & {\bf B} & {\bf A}_{2} & \ldots & 0 \\
\vdots & \vdots & \vdots & \vdots & \vdots \\
0 & 0 & 0 & \ldots & {\bf A}_{S+1}
\end{array} \right) \left( \begin{array}{c}
{\bf C}_{0} \\ {\bf C}_{1} \\ {\bf C}_{2} \\ \vdots \\ {\bf
C}_{S+1} \end{array} \right) = \left( \begin{array}{c} {\bf D} \\
0 \\ 0 \\ \vdots \\ 0 \end{array} \right) ,
\end{eqnarray}
with
\begin{eqnarray}\label{c02}
{\bf A}_{0} &=& E{\bf I} - \widetilde{{\bf H}}_0 \nonumber \\
{\bf A}_{i} &=& E{\bf I} - {\bf H}_i \,,\;i=1,\ldots,S \nonumber
\\
{\bf A}_{S+1} &=& E{\bf I} - \widetilde{{\bf H}}_{S+1} \nonumber \\
{\bf C}_{0} &=& {\bf U}_{\rm L}(+) + {\bf U}_{\rm L}(-){\bf R}
\nonumber \\
{\bf C}_{S+1} &=& {\bf U}_{\rm R}(+){\bf T} \nonumber \\
{\bf D} &=& {\bf Q}{\bf U}_{\rm L}(+) .
\end{eqnarray}
All the blocks ${\bf A}$-${\bf D}$ are $N_{\rm rs} \times N_{\rm
rs}$ matrices. Eq.~(\ref{c01}) represents a set of linear
equations, which can be solved directly using a standard
algorithm. However, the dimension of this problem is $N_{\rm
tot}=N_{\rm rs} \cdot (S+2)$, which can be rather large. Since the
computing cost scales as $N_{\rm tot}^3$ the direct route is not
very practical.

It is however quite straightforward to construct an algorithm for which the computing cost
scales as $N_{\rm rs}^3 \cdot S$, i.e. only linearly with the size $S$ of the scattering
region. One has to make optimal use of the block tridiagonal form of the matrix in
Eq.~(\ref{c01}). The algorithm is a block form of Gaussian elimination. The first (and most
time consuming) step of this algorithm is transforming the matrix into upper block triangular
form by iteration
\begin{eqnarray}\label{c03}
&&{\bf A'}_{0}={\bf A}_{0}\,;\; {\bf D'}_{0}={\bf D}\,; \nonumber \\
&& \left. \begin{array}{ccc} {\bf A'}_{i}={\bf A}_{i} - {\bf
B}{\bf A'}_{i-1}^{-1}{\bf
B^{\dagger}} \\
{\bf D'}_{i}=-{\bf B}{\bf A'}_{i-1}^{-1}{\bf D'}_{i-1}
\end{array} \right\} i=1,\ldots,S+1 .
\end{eqnarray}
The inverse matrices ${\bf A'}_{i-1}^{-1}$ in this algorithm are actually not needed
explicitly. Instead at each step one solves the sets of linear equations
\begin{equation}\label{c03a}
{\bf A'}_{i-1}\widetilde{{\bf B}}_{i}={\bf B^{\dagger}}\,;\;{\bf
A'}_{i-1}\widetilde{{\bf D}}_{i}={\bf D'}_{i-1} ,
\end{equation}
by a standard algorithm, i.e. $LU$ decomposition of ${\bf A'}_{i-1}$ followed by back
substitution, to obtain the matrices $\widetilde{{\bf B}}_{i}$ and $\widetilde{{\bf
D}}_{i}$.\cite{golub} This allows the steps in Eq.~(\ref{c03}) to be rewritten as
\begin{equation}\label{c03b}
{\bf A'}_{i}={\bf A}_{i} - {\bf B}\widetilde{{\bf B}}_{i}\,;\;
{\bf D'}_{i}=-{\bf B}\widetilde{{\bf D}}_{i} .
\end{equation}

The solution to Eq.~(\ref{c01}) can now be found by back
substitution
\begin{eqnarray}\label{c04}
&& {\bf C}_{S+1}={\bf A'}_{S+1}^{-1}{\bf D'}_{S+1}  \nonumber \\
&& \left. {\bf C}_{i}=\widetilde{{\bf D}}_{i+1} - \widetilde{{\bf
B}}_{i+1}{\bf C}_{i+1} \right\} i=S,\ldots,0 .
\end{eqnarray}
Again one does not need ${\bf A'}_{S+1}^{-1}$ explicitly, but like
Eq.~(\ref{c03a}) one can solve the equivalent set of linear
equations. The reflection and transmission matrices ${\bf R}$ and
${\bf T}$ can be extracted using the special form of the matrices
${\bf C}_{0}$ and ${\bf C}_{S+1}$, see Eq.~(\ref{c02}).

Very often one is only interested in the transmission matrix. In
that case one only uses the first step of the back substitution,
Eq.~(\ref{c04}), which can be written as
\begin{equation}\label{c05}
{\bf A'}_{S+1}{\bf U}_{\rm R}(+){\bf T}={\bf D'}_{S+1} .
\end{equation}
This is a set of linear equations for the transmission probability amplitudes ${\bf T}$,
which can be solved using a standard numerical techniques.\cite{golub}

The time consuming steps consist of solving Eq.~(\ref{c03a}), the
computing costs of which scale as $N_{\rm
rs}^3$.\cite{speedupremark} Using Eq.~({\ref{c03b}) in
Eq.~(\ref{c03}) requires performing $S+1$ of such steps and
subsequently solving Eq.~(\ref{c05}) scales as $N_{\rm rs}^3$.
Note that the full algorithm scales linearly with the size $S$ of
the scattering region.

\subsection{Velocities}\label{velocities}
In this section we give a short derivation of the expression for the velocities,
Eq.~(\ref{f018}). It is straightforward to show that the vectors ${\bf u}_m$ of
Eq.~(\ref{f03a}) are a solution of the quadratic eigenvalue equation
\begin{equation}\label{d01}
\lambda_m (E{\bf I} - {\bf H}) {\bf u}_m + {\bf B}{\bf u}_m +
\lambda_m^2 {\bf B^{\dagger}} {\bf u}_m = 0.
\end{equation}
This quadratic eigenvalue equation of dimension $N_{\rm rs}$ is
completely equivalent to the linear problem of dimension $2N_{\rm
rs}$ of Eq.~(\ref{f7}). If ${\bf u}_m$ is a right eigenvector of
Eq.~(\ref{d01}) belonging to the eigenvalue $\lambda_m$, then by
complex conjugation of this equation one shows that ${\bf
u}_m^\dagger$ is a left eigenvector belonging to the eigenvalue
$1/\lambda_m^*$. For a propagating state $|\lambda_m|=1$, so
$\lambda_m = 1/\lambda_m^*$, which means that these left and right
eigenvectors belong to the same eigenvalue.

We now start from
\begin{equation}\label{d02}
\lambda_m {\bf u}_m^\dagger (E{\bf I} - {\bf H}) {\bf u}_m + {\bf
u}_m^\dagger {\bf B}{\bf u}_m + \lambda_m^2 {\bf u}_m^\dagger {\bf
B^{\dagger}} {\bf u}_m = 0,
\end{equation}
and take the derivative $d/dE$ of this expression. All the terms
with $d{\bf u}_m/dE$ and $d{\bf u}_m^\dagger/dE$ drop out, because
${\bf u}_m$ and ${\bf u}_m^\dagger$ obey Eq.~(\ref{d01}) and its
complex conjugate, respectively. The remaining terms can be
collected and slightly rewritten using Eq.~(\ref{d01}); the result
is
\begin{eqnarray}\label{d03}
&& \frac{d\lambda_m}{dE} \left( \lambda_m^{-1} {\bf u}_m^\dagger
{\bf B}{\bf u}_m - \lambda_m {\bf u}_m^\dagger {\bf B^{\dagger}}
{\bf u}_m \right) + \lambda_m {\bf u}_m^\dagger {\bf u}_m =
\nonumber \\
&& -2 i \frac{d\lambda_m}{dE} {\rm Im} \left( \lambda_m {\bf
u}_m^\dagger {\bf B^{\dagger}} {\bf u}_m \right) + \lambda_m = 0,
\end{eqnarray}
where the last line is obtained by making use of
$\lambda_m^{-1}=\lambda_m^*$ and the fact that the vectors are
normalized, ${\bf u}_m^\dagger {\bf u}_m = 1$. Eq.~(\ref{d03})
yields an expression for $d\lambda_m/dE$. For propagating states
$\lambda_m = e^{i k_x a_x}$ and thus
\begin{equation}\label{d04}
\frac{dk_x}{dE} = \frac{1}{i a_x \lambda_m} \frac{d\lambda_m}{dE}
\end{equation}
The usual definition of the Bloch velocity $v_n=\hbar^{-1}
dE/dk_x$ and the expression for $d\lambda_m/dE$ extracted from
Eq.~(\ref{d03}) then give the expression for the velocity of
Eq.~(\ref{f018}).


\begin{thebibliography}{999}
\bibitem{greenremark} If one uses Green functions, one has to
distinguish between retarded and advanced Green functions. This
can be done by subtracting or adding a (small) imaginary number to
the energy.
\bibitem{cells} The size of the scattering region does not have to be an
integer multiple of the size of the cell the leads. One can choose
cells of a different size in the scattering region. One also
chooses differently sized cells in the left and right leads in
case the leads are different. We have not introduced cells of
different size here in order not to complicate the notation, but
the strategy for solving the scattering problem as outlined in
appendix \ref{scatt} also works in that case.
\bibitem{dimremark}
The number of cells $S$ required for the scattering region of a $n$-atomic wire is
approximately $n_{\mathrm L}+n_{\mathrm R}+n$, where $n_{\mathrm L},n_{\mathrm R}$ is the
number of unit cells that define the surface regions of left and right electrodes. The number
of grid points per cell is $N_{\mathrm rs}=L\cdot W_{y}\cdot W_{z}=3456$.
\bibitem{potremark} Somewhat surprisingly, the size of the even-odd oscillation is however
quite sensitive to the accuracy of the sodium potential used as input for the transport
calculations. Extracting this potential from a self-consistent ground state calculation
requires the ground state energy to be converged at the scale of $10^{-6}$-$10^{-7}$
Hartrees. A weaker convergence criterion of $10^{-5}$ Hartrees yields a potential that can
change the conductance by up to 5 \% and can even suppress the even-odd oscillation almost
completely.
\end{thebibliography}
\end{document}